\documentclass[12pt]{article}
\usepackage{amscd}
\usepackage{bbm}
\usepackage{mathrsfs}
\usepackage{amssymb}
\usepackage{graphics}
\usepackage{graphicx}
\usepackage{amsfonts}
\usepackage{amsmath}
\usepackage{float}
\usepackage[numbers,sort&compress]{natbib}

\begin{document}
\date{}
\title{Stability of elliptic solutions to the sinh-Gordon equation}
\author{ Wen-Rong Sun$^{1,2}$ and Bernard Deconinck$^{2}$ \\ \\
1. School of Mathematics and Physics,\\ University
of Science and Technology Beijing, Beijing 100083, China\\
2. Department of Applied Mathematics, \\
University of Washington, Seattle, WA 98195, USA}
\maketitle \vspace{-0.8cm}


\begin{abstract}
Using the integrability of the sinh-Gordon equation, we demonstrate the spectral
stability of its elliptic solutions. By
constructing a Lyapunov functional using higher-order conserved quantities of the sinh-Gordon equation, we show that these elliptic solutions are orbitally stable with respect to subharmonic perturbations of arbitrary period.
\textbf{AMS subject classications.} 37K45, 35Q58, 33E05
\end{abstract}
\newpage

\noindent\textbf{\large 1. Introduction}\\\hspace*{\parindent}
In real space-time coordinates, denoted $(X, T)$,
the sinh-Gordon equation reads~\cite{sinh1,sinh2}
\begin{eqnarray}\label{e2}
u_{TT}-u_{XX}+\sinh (u)=0,
\end{eqnarray}
where partial derivatives are denoted by subscripts. Passing to light-cone coordinates $(x,t)$, with
\begin{eqnarray}\label{tr1}
x=\frac{1}{2}(X+T), \quad t=\frac{1}{2}(X-T),
\end{eqnarray}
the sinh-Gordon equation becomes
\begin{equation}\label{e1}
u_{x t}= \sinh u.
\end{equation}
The sinh-Gordon equation~(\ref{e2})  is rewritten in system form as
\begin{subequations}\label{e31}
\begin{eqnarray}
&& u_{T}=-p,\\
&& p_{T}+u_{XX}-\sinh (u)=0.
\end{eqnarray}
\end{subequations}
Here $u$ is a real-valued function. The sinh-Gordon equation arises in the context of particular
surfaces of constant mean curvature. The geometrical interpretation of (\ref{e2}) was shown by
studying surfaces of constant Gaussian curvature in a three-dimensional pseudo-Riemannian
manifold of constant curvature~\cite{sinh3}. It has appeared in differential geometry and various
applications. For example, it can be used to describe generic properties of string dynamics
for strings and multi-strings in constant curvature space~\cite{sinh4}.  Equation~(\ref{e1}) is an integrable system and has a self-adjoint Lax pair~\cite{sinh1}. The stability of periodic wave solutions was
studied in~\cite{sinh5} where it was shown that the periodic wave solutions are orbitally stable for $(u,p)\in
H_{m, p e r}^{1}([0, L]) \times L_{p e r}^{2}([0, L])$, for specific choices of the traveling wave velocity.
Here, $H_{m, p e r}^{1}([0, L])=\left\{f \in H_{p e r}^{1}([0, L]) ;\frac{1}{L} \int_{0}^{L} f(x) d x=0\right\}$.

In this paper, using the integrability of~(\ref{e2}), we show the spectral and orbital stability
of elliptic solutions of the sinh-Gordon equation with respect to arbitrary subharmonic perturbations:
perturbations that are periodic with period equal to an integer multiple of the
period of the underlying solution.
 Using the same method as in~\cite{bd5,bd6,bd7,bd8,t1}, we prove the spectral
stability of the elliptic solutions by providing an explicit analytic description of the spectrum
and the eigenfunctions associated with the linear stability problem of the elliptic solutions. Next, as in~\cite{bd4,bd5,bd6,bd7,by90}, we employ the Lyapunov method~\cite{tt11,tt22}, which has formed the crux of
subsequent nonlinear stability techniques~\cite{lf1,lf2,lf3}, to conclude the orbital stability with the help of classical results of Grillakis, Shatah and Strauss~\cite{lf4}.
\\
\\
\noindent\textbf{\large 2. The elliptic solutions of the sinh-Gordon equation}\\\hspace*{\parindent}
We construct the real, bounded, periodic, traveling wave solutions to the sinh-Gordon
equation. To obtain the traveling wave solutions, one rewrites the sinh-Gordon equation in a frame moving with constant velocity $c$. With $z=X-cT$ and $\tau=T$, the sinh-Gordon equation becomes
\begin{eqnarray}\label{b2}
\left(c^{2}-1\right) u_{z z}-2 c u_{z \tau}+u_{\tau \tau}+\sinh (u)=0.
\end{eqnarray}
In what follows, we assume that $c \neq \pm 1$.
Stationary solutions are time-independent solutions of~(\ref{b2}).
They satisfy the ordinary differential equation
\begin{eqnarray}\label{ds}
\left(c^{2}-1\right) f^{\prime \prime}(z)+\sinh (f(z))=0, \quad^{\prime} :=\frac{d}{d z}.
\end{eqnarray}
Multiplying by $f'(z)$ and integrating once,
\begin{eqnarray}\label{cd1}
\frac{1}{2}\left(c^{2}-1\right) f^{\prime}(z)^{2}+\cosh (f(z))=\mathcal{E},
\end{eqnarray}
where $\mathcal{E}$ is a constant of integration referred to as the total energy.

Equation~(\ref{ds}) is rewritten as the first-order two-dimensional system
\begin{equation}
f^{\prime}(z)=g(z), \quad g^{\prime}(z)=\frac{-2\sinh(f(z))}{c^2-1},
\end{equation}
with $(0,0)$ as a fixed point. The linearization about the origin has eigenvalues
\begin{equation}
\lambda=\pm \sqrt{\frac{-2}{c^2-1}}.
\end{equation}
Thus small-amplitude periodic solutions are expected for
$c^2-1>0$. If $c^2>1$ then $\mathcal{E}>1$, from~(\ref{cd1}).

Motivated by~\cite{sinh5}, we look for
solutions to~(\ref{cd1}) of the form
\begin{eqnarray}\label{scc1}
\cosh(f(z))=\frac{\alpha}{1-\beta v(z)^2}+d,
\end{eqnarray}
where $v(z)$ is a function to be determined, and $\alpha$, $\beta$ and $d$ are parameters.
Differentiating and squaring ~(\ref{scc1}), one obtains
\begin{eqnarray}\label{sd1}
\frac{4\alpha^2 \beta^2 v^2}{(1-\beta v^2)^4}\left(\frac{d v}{d z}\right)^{2}=\sinh^2(f(z))\left(\frac{d f}{d z}\right)^{2}.
\end{eqnarray}
Using~(\ref{cd1}), the above equation can be reduced to
\begin{eqnarray}\label{sss2}
\left(\frac{d v}{d z}\right)^{2}=\frac{[\alpha^2(1-\beta v^2)+2\alpha d (1-\beta v^2)^2+(d^2-1)(1-\beta v^2)^3][2\mathcal{E}-2d-2\alpha-(2\mathcal{E}-2d)\beta v^2]}{4\alpha^2 \beta^2v^2(c^2-1)}.
\end{eqnarray}
Here $\operatorname{sn}(z, k)$ denotes the Jacobi elliptic sine function with argument $z$ and
modulus $k$~\cite{rb1}, which satisfies the first-order nonlinear equation
\begin{eqnarray}\label{sn1}
\left(\frac{d v}{d z}\right)^{2}=\left(1-v^{2}\right)\left(1-k^{2} v^{2}\right).
\end{eqnarray}
Motivated by (\ref{sn1}), we wish to eliminate the higher-order terms in the numerator and $v^2$ from the denominator of (\ref{sss2}). This is accomplished by equating
$d^2=1$ and $\alpha+2d=0$ or $d^2=1$ and $\mathcal{E}=d+\alpha$.

\textbf{\emph{Case I: }} with the condition $d^2=1$ and $\mathcal{E}=d+\alpha$, we cannot find elliptic solutions from (\ref{sss2}). In fact, with $d^2=1$ and $\mathcal{E}=d+\alpha$, motivated by the form of (\ref{sn1}),  the expression
of~(\ref{sss2}) implies that $\beta=1$ or $\beta=1+\frac{\alpha}{2d}$.

$\bullet$ For  $\beta=1+\frac{\alpha}{2d}$, Equation~(\ref{sss2}) becomes
\begin{eqnarray}\label{rb2}
\left(\frac{d v}{d z}\right)^{2}=\frac{(1-\beta v^2)[2\alpha d \beta(1- v^2)](-2\mathcal{E}+2d)\beta}{4\alpha^2 \beta^2(c^2-1)}.
\end{eqnarray}
Comparing (\ref{rb2}) and (\ref{sn1}), we obtain $0<\beta<1$ since the elliptic modulus $0<k<1$ in (\ref{sn1}).  Since $d^2=1$ and $\mathcal{E}>1$, we know that $\alpha=\mathcal{E}-d>0$. From $\beta=1+\frac{\alpha}{2d}$, we know $d=-1$ so that $0<\beta<1$. From $\beta=1-\frac{\alpha}{2}>0$, we know $\alpha<2$. But when $d=-1$, we know that $\alpha=\mathcal{E}+1>2$.

$\bullet$ For $\beta=1$ and $d=1$, Equation~(\ref{sss2}) becomes
\begin{eqnarray}
\left(\frac{d v}{d z}\right)^{2}=\frac{\left(1-v^{2}\right)\left(\alpha^{2}+2 \alpha\right)\left[\left(1-\frac{2 \alpha}{\alpha^{2}+2 \alpha} v^{2}\right)\right](-2 \mathcal{E}+2)}{4 \alpha^{2}\left(c^{2}-1\right)},
\end{eqnarray}
from which $\frac{(-2\mathcal{E}+2)(\alpha^2+2\alpha)}{4\alpha^2(c^2-1)}<0$, and no elliptic solutions are obtained from~(\ref{sn1}).

$\bullet$ For $\beta=1$ and $d=-1$,  Equation~(\ref{sss2}) becomes
\begin{eqnarray}
\left(\frac{d v}{d z}\right)^{2}=\frac{\left(1-v^{2}\right)\left(\alpha^{2}-2 \alpha\right)\left[\left(1+\frac{2 \alpha}{\alpha^{2}-2 \alpha} v^{2}\right)\right](-2 \mathcal{E}-2)}{4 \alpha^{2}\left(c^{2}-1\right)}.
\end{eqnarray}
Since $d=-1$, $\alpha=\mathcal{E}-d>2$, which implies that $-\frac{2 \alpha}{\alpha^{2}-2 \alpha}<0$. Therefore we cannot obtain
elliptic solutions from~(\ref{sn1}).

\textbf{\emph{Case II: }} we consider $d^2=1$ and $\alpha+2d=0$.

Equation~(\ref{sss2}) is reduced to
\begin{eqnarray}\label{sss222}
\left(\frac{d v}{d z}\right)^{2}=\frac{(1-\beta v^2)[2\mathcal{E}-2d-2\alpha-(2\mathcal{E}-2d)\beta v^2]}{4 \beta(c^2-1)},
\end{eqnarray}

so that

\begin{eqnarray}
v=\operatorname{sn}(b z, k), k=\sqrt{\frac{\mathcal{E}-1}{\mathcal{E}+1}}, \alpha=2, \beta=k^2, d=-1, b=\sqrt{\frac{\mathcal{E}+1}{2(c^2-1)}}.
\end{eqnarray}

It follows that

\begin{eqnarray}\label{sc1}
\cosh(f(z))=\frac{2}{1-k^2 \operatorname{sn}(b z, k)^2}-1.
\end{eqnarray}
The solution is periodic with  period
$T(k)=\frac{2 \mathrm{K}}{b}$, where

\begin{equation}
K(k)=\int_{0}^{\pi / 2} \frac{\mathrm{d} y}{\sqrt{1-k^{2} \sin ^{2}(y)}},
\end{equation}
the complete elliptic integral of the first kind, see~\cite{rb1}.
\\
\\
\noindent\textbf{\large 3. The linear stability problem}\\\hspace*{\parindent}
In this section, we examine the stability of the elliptic solutions obtained above.
Considering the perturbation of a stationary solution to (\ref{b2}),
\begin{eqnarray}
u(z, \tau)=f(z)+\epsilon w(z, \tau)+\mathcal{O}\left(\epsilon^{2}\right),
\end{eqnarray}
where $\epsilon$ is a small parameter, we obtain the linear stability problem
\begin{eqnarray}\label{B1}
\left(c^{2}-1\right) w_{z z}-2 c w_{z \tau}+w_{\tau \tau}+\cosh (f(z)) w=0.
\end{eqnarray}
With $w_{1}(z, \tau)=w(z, \tau) \text { and } w_{2}(z, \tau)=c w_{z}(z, \tau)-w_{\tau}(z, \tau)$, the linear problem is rewritten as
\begin{equation}\label{asd}
\frac{\partial}{\partial \tau}\left(\begin{array}{c}{w_{1}} \\ {w_{2}}\end{array}\right)=\left(\begin{array}{cc}{c \partial_{z}} & {-1} \\ {- \partial_{z}^{2}+\cosh (f(z))} & {c \partial_{z}}\end{array}\right)\left(\begin{array}{c}{w_{1}} \\ {w_{2}}\end{array}\right).
\end{equation}
We note that (\ref{asd}) is autonomous in time.
By separating variables,
\begin{eqnarray}\label{fg1}
\left(\begin{array}{c}{w_{1}(z, \tau)} \\ {w_{2}(z, \tau)}\end{array}\right)=e^{\lambda \tau}\left(\begin{array}{c}{W_{1}(z)} \\ {W_{2}(z)}\end{array}\right),
\end{eqnarray}
the linear problem~(\ref{asd}) is rewritten as

\begin{eqnarray}\label{sqj1}
\lambda\left(\begin{array}{c}{W_{1}} \\ {W_{2}}\end{array}\right)=J\mathcal{L}\left(\begin{array}{c}{W_{1}} \\ {W_{2}}\end{array}\right)=\left(\begin{array}{cc}{c \partial_{z}} & {-1} \\ {- \partial_{z}^{2}+\cosh (f(z))} & { c \partial_{z}}\end{array}\right)\left(\begin{array}{c}{W_{1}} \\ {W_{2}}\end{array}\right),
\end{eqnarray}
where

\begin{eqnarray}
J=\left(\begin{array}{cc}{0} & {1} \\ {-1} & {0}\end{array}\right), \mathcal{L}=\left(\begin{array}{ccc} { \partial_{z}^{2}-\cosh (f(z))} & {- c \partial_{z}}\\ c \partial_{z} & -1 \end{array}\right).
\end{eqnarray}
Note that ${\mathcal{L}}$ is formally self adjoint. We define
\begin{equation}
\sigma_{J\mathcal{L}}=\left\{\lambda \in \mathbb{C} : \sup _{x \in \mathbb{R}}\left(\left|W_{1}(x)\right|,\left|W_{2}(x)\right|\right)<\infty\right\}.
\end{equation}

Spectral stability of an elliptic solution with respect to perturbations that are bounded on the whole line is established by
demonstrating that the spectrum $\sigma(J \mathcal{L})$ of the operator $J \mathcal{L}$ does not intersect the right-half complex $\lambda$ plane. Because the sinh-Gordon equation is a Hamiltonian partial differential equation~\cite{sinh2}, the spectrum is
symmetric under reflection with respect to both the real and imaginary axes~\cite{bbby9}. As a consequence, spectral stability requires that $\sigma(J \mathcal{L})$ is confined to the imaginary axis.
\\
\\
\noindent\textbf{\large 4. The Lax pair restricted to the elliptic solution}\\\hspace*{\parindent}
Equation~(\ref{e1}) admits the following Lax pair~\cite{sinh1}:
\begin{eqnarray}\label{T1}
\psi_{x}=\left( \begin{array}{cc}{-i \zeta} & {\frac{u_{x}}{2}} \\ {\frac{u_{x}}{2}} & {i \zeta}\end{array}\right) \psi,\quad
\psi_{t}=\left( \begin{array}{cc} \frac{i\cosh u}{4\zeta} & -\frac{i\sinh u}{4\zeta} \\ \frac{i\sinh u}{4\zeta}  & -\frac{i\cosh u}{4\zeta}\end{array}\right) \psi.
\end{eqnarray}
From (\ref{T1}), one has the following spectral problem:
\begin{equation}\label{t3}
\left( \begin{array}{cc}{i \partial_{x}} & {-\frac{i}{2} u_{x}} \\ {\frac{i}{2} u_{x}} & {-i \partial_{x}}\end{array}\right) \psi=\zeta \psi.
\end{equation}
Since the spectral problem is self adjoint, the Lax spectrum $\sigma_L$ is a subset of the real line: $\sigma_L:= \{\zeta\in \mathbb{C}: \sup _{x \in \mathbb{R}}{(|\psi_1|,|\psi_2|)}<\infty\}\subset \mathbb{R}$.

Through (\ref{tr1}), (\ref{e2}) admits the following Lax pair:
\begin{eqnarray}\label{T1r1}
&&\psi_{X}=\left( \begin{array}{cc}{-\frac{i}{2} \zeta+\frac{i\cosh u}{8\zeta}} & {\frac{u_{X}+u_{T}}{4}-\frac{i\sinh u}{8\zeta}} \\ {\frac{u_{X}+u_{T}}{4}+\frac{i\sinh u}{8\zeta}} & {\frac{i}{2} \zeta-\frac{i\cosh u}{8\zeta}}\end{array}\right) \psi,\nonumber\\
&&\psi_{T}=\left( \begin{array}{cc}{-\frac{i}{2} \zeta-\frac{i\cosh u}{8\zeta}} & {\frac{u_{X}+u_{T}}{4}+\frac{i\sinh u}{8\zeta}} \\ {\frac{u_{X}+u_{T}}{4}-\frac{i\sinh u}{8\zeta}} & {\frac{i}{2} \zeta+\frac{i\cosh u}{8\zeta}}\end{array}\right) \psi.\nonumber
\end{eqnarray}
We transform the Lax pair by moving into a traveling reference frame,  letting $z=X-c T$, $\tau=T$ and $u(z,\tau)=f(z)$. The Lax pair restricted to the stationary solution is
\begin{eqnarray}\label{T1r2}
&&\psi_{z}=\left( \begin{array}{cc}{-\frac{i}{2} \zeta+\frac{i\cosh f(z)}{8\zeta}} & {\frac{(1-c)f'(z)}{4}-\frac{i\sinh f(z)}{8\zeta}} \\ {\frac{(1-c)f'(z)}{4}+\frac{i\sinh f(z)}{8\zeta}} & {\frac{i}{2} \zeta-\frac{i\cosh f(z)}{8\zeta}}\end{array}\right) \psi,\\
&&\psi_{\tau}=\left(\begin{array}{cc}{A} & {B} \\ {C} & {-A}\end{array}\right) \psi=\left( \begin{array}{cc}{-\frac{i(1+c)}{2} \zeta+(c-1)\frac{i\cosh f(z)}{8\zeta}} & {\frac{(1-c)(c+1)f'(z)}{4}+(1-c)\frac{i\sinh f(z)}{8\zeta}} \\ {\frac{(1-c)(c+1)f'(z)}{4}+(c-1)\frac{i\sinh f(z)}{8\zeta}} & {(c+1)\frac{i}{2} \zeta+(1-c)\frac{i\cosh f(z)}{8\zeta}}\end{array}\right) \psi.\nonumber
\end{eqnarray}
Since $A$, $B$ and $C$ are independent of $\tau$, we separate variables to look for solutions of the form
\begin{eqnarray}\label{ssd}
\psi(z, \tau)=e^{\Omega \tau} \varphi(z),
\end{eqnarray}
where $\Omega$ is independent of $\tau$. We substitute (\ref{ssd}) into the $\tau$-part of the Lax pair and obtain
\begin{equation}
\left(\begin{array}{cc}{A-\Omega} & {B} \\ {C} & {-A-\Omega}\end{array}\right) \varphi=0.
\end{equation}
This implies that the existence of nontrivial solutions requires
\begin{eqnarray}\label{tsd}
\Omega^{2}=A^{2}+BC=-\frac{16 \zeta^4 (c+1)^2-8 \zeta^2 \left(c^2-1\right) \mathcal{E}+(c-1)^2}{64 \zeta^2},
\end{eqnarray}
where we have used the explicit form of the stationary solution ${f'(z)}^2$ derived
earlier. Equation~(\ref{tsd}) determines $\Omega$ in terms of $\zeta$. Since $\zeta$ is real, $\Omega$ is real or imaginary. Further, $\Omega^2$ is an even function of $\zeta$. From the discriminant of (\ref{tsd}), we know that with $\mathcal{E}>1$ and $c^2>1$, (\ref{tsd}) is expressed as
\begin{eqnarray}\label{tsd2}
\Omega^{2}=-\frac{(\zeta-\zeta_{1})(\zeta-\zeta_{2})(\zeta+\zeta_{2})(\zeta+\zeta_{1})}{4 \zeta^2},
\end{eqnarray}
where $\zeta_{1}=\frac{1}{2} \sqrt{\frac{\left(c-1\right) (\mathcal{E}-\sqrt{ \mathcal{E}^2-1})}{(c+1)}}$ and $\zeta_{2}=\frac{1}{2} \sqrt{\frac{\left(c-1\right) (\mathcal{E}+\sqrt{ \mathcal{E}^2-1})}{(c+1)}}$, see Figure~1.

\begin{center}
{\includegraphics[scale=0.90]
{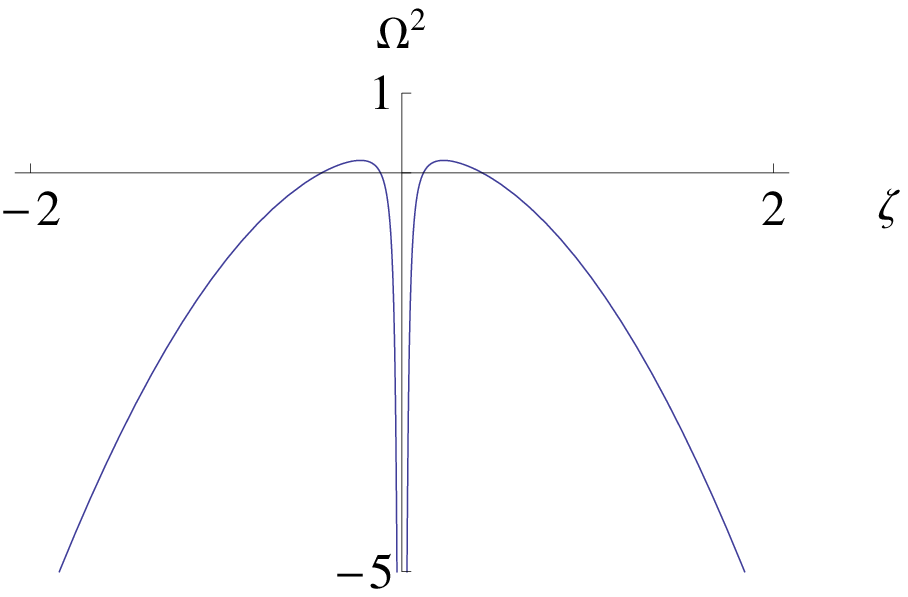}}
\\\vspace{-0.1cm}{\footnotesize {\bf Figure~1}\
The graph of $\Omega^2$ as a function of real $\zeta$.}
\end{center}

The eigenvectors corresponding to the eigenvalue $\Omega$ are
\begin{eqnarray}\label{ef1}
\varphi(z)=\gamma(z)\left(\begin{array}{c}{-B(z)} \\ {A(z)-\Omega}\end{array}\right),
\end{eqnarray}
where $\gamma(z)$ is a scalar function. It is determined by substitution of (\ref{ef1})
into the $z$-part of the Lax pair, resulting in a first-order scalar differential
equation for $\gamma(z)$, so that
\begin{eqnarray}\label{sim1}
\gamma(z)&=&\exp \left[\int \left(\frac{i}{2} \zeta-\frac{i\cosh f(z)}{8\zeta} - \frac{B \left[{\frac{(1-c)f'(z)}{4}+\frac{i\sinh f(z)}{8\zeta}}\right]+A'}{A-\Omega}\right) \mathrm{d} z\right]\nonumber\\
&=&\frac{1}{A-\Omega} \exp \left[\int \left(\frac{i}{2} \zeta-\frac{i\cosh f(z)}{8\zeta} - \frac{B \left[{\frac{(1-c)f'(z)}{4}+\frac{i\sinh f(z)}{8\zeta}}\right]}{A-\Omega}\right) \mathrm{d} z\right].
\end{eqnarray}
Excluding the branch points, where $\Omega=0$, each $\zeta$ results in two values of $\Omega$. Thus (\ref{ef1}) represents two linearly independent solutions. When $\zeta=\zeta_{1}$ or $\zeta=\zeta_{2}$, only one solution is obtained. The second one may be obtained using the reduction of order method.

Since the vector part of the eigenvector $\varphi(z)$ is bounded in $z$, we need to check for which $\zeta$ the scalar function $\gamma$ is bounded for all $z$, including as $|z|\rightarrow \infty$. A necessary and sufficient condition for this is that~\cite{bd6,bd8}
\begin{eqnarray}\label{con1}
\left\langle\operatorname{Re}\left(- \frac{B \left[{\frac{(1-c)f'(z)}{4}+\frac{i\sinh f(z)}{8\zeta}}\right]+A'}{A-\Omega}\right)\right\rangle= 0.
\end{eqnarray}
Here $\langle\cdot\rangle=\frac{1}{ T} \int_{-\frac{T}{2}}^{\frac{T}{2}} \cdot d z$ denotes the average over a period. The explicit form of the above depends on $\Omega$. Since $\zeta$ is real, $\Omega$ is real or imaginary from (\ref{tsd}). Recently, Upsal and Deconinck~\cite{jm} demonstrated that purely real Lax spectrum implies spectral stability. We show this explicitly below.

a) When $\Omega$ is imaginary, the integrand of (\ref{con1}) is
\begin{eqnarray}
\operatorname{Re}\left(- \frac{B \left[{\frac{(1-c)f'(z)}{4}+\frac{i\sinh f(z)}{8\zeta}}\right]}{A-\Omega}\right)+\operatorname{Re}\left(- \frac{A'}{A-\Omega}\right).
\end{eqnarray}
The second term is a total derivative,
with zero average over a period.
For the first term,
\begin{eqnarray}
\operatorname{Re}\left(- \frac{B \left[{\frac{(1-c)f'(z)}{4}+\frac{i\sinh f(z)}{8\zeta}}\right]}{A-\Omega}\right)=f'(z)\left[\frac{\sinh f(z)(1-c^2)+\sinh f(z)(1-c)^2}{32\zeta(-i A+i\Omega)}\right],
\end{eqnarray}
which is a  total derivative,
resulting in zero average over a period. Therefore, all $\zeta$ values for which $\Omega$
is imaginary are in the Lax spectrum.

b) If $\Omega$ is real, the second term is still a total derivative,
thus giving zero average over a period. We consider the first term:
\begin{eqnarray}
\operatorname{Re}\left(- \frac{B \left[{\frac{(1-c)f'(z)}{4}+\frac{i\sinh f(z)}{8\zeta}}\right]}{A-\Omega}\right)=\Omega\frac{\frac{(1+c)(1-c)^2 f'(z)^2}{16}+\frac{(c-1)(\sinh^2(f(z)))}{64\zeta^2}}{\Omega^2+A^2}+f'(z)F(f(z)).
\end{eqnarray}
The second term is a total derivative,
thus giving zero average over a period. The first term results in a zero average only when $\Omega$ is zero. Thus all values of $\zeta$ for which
 $\Omega$ is real (except for the four branch points, where $\Omega=0$) are not part of the
Lax spectrum.

We conclude that the Lax spectrum consists of all $\zeta$ values for which $\Omega^{2} \leq 0$:
\begin{equation}
\sigma_{L}=\left(-\infty,-\zeta_{2}\right] \cup\left[-\zeta_{1},0\right) \cup\left(0, \zeta_{1}\right] \cup\left[\zeta_{2}, \infty\right).
\end{equation}
Moreover, $\Omega^2$ takes on all negative values for $\left(-\infty,-\zeta_{2}\right]$, implying that $\Omega$ covers the imaginary axis. The same is true for the other intervals. Thus
\begin{eqnarray}\label{tfw}
\Omega \in(i \mathbb{R})^{4}, \quad \zeta\in \sigma_L,
\end{eqnarray}
where the exponent denotes multiplicity.
\\
\\
\noindent\textbf{\large 5. The squared eigenfunction connection}\\\hspace*{\parindent}
It is well known that there exists a connection between the eigenfunctions of
the Lax pair of an integrable equation and the eigenfunctions of the linear
stability problem for this integrable equation~\cite{sinh1,by11,by12,bd5,bd6,bd7,bd8,t1}.

\textbf{Theorem 1} \emph{The difference of squares,
\begin{eqnarray}\label{www1}
w(z, \tau)=\psi_{1}(z, \tau)^{2}-\psi_{2}(z, \tau)^{2},
\end{eqnarray}
satisfies the linear stability problem~(\ref{B1}). Here $\psi=\left(\psi_{1}, \psi_{2}\right)^{T}$ is any solution of the Lax pair~(\ref{T1r2}).}

To establish the connection between the $\sigma_{J\mathcal{L}}$ spectrum and the $\sigma_{L}$ spectrum we examine the
right- and left-hand sides of (\ref{fg1}). Substituting (\ref{www1}) and (\ref{ssd}) into (\ref{fg1}),
\begin{equation}
e^{2 \Omega \tau}\left(\begin{array}{c}{\varphi_{1}^{2}-\varphi_{2}^{2}} \\ {-2 \Omega\left(\varphi_{1}^{2}-\varphi_{2}^{2}\right)+2c \left(\varphi_{1}\varphi_{1z}-\varphi_{2}\varphi_{2z}\right)}\end{array}\right)=e^{\lambda \tau}\left(\begin{array}{c}{W_{1}(z)} \\ {W_{2}(z)}\end{array}\right),
\end{equation}
so that
\begin{equation}
\lambda=2 \Omega(\zeta),
\end{equation}
with
\begin{eqnarray}\label{tsj1}
\left(\begin{array}{c}{W_{1}(z)} \\ {W_{2}(z)}\end{array}\right)=\left(\begin{array}{c}{\varphi_{1}^{2}-\varphi_{2}^{2}} \\ {-2 \Omega\left(\varphi_{1}^{2}-\varphi_{2}^{2}\right)+2c \left(\varphi_{1}\varphi_{1z}-\varphi_{2}\varphi_{2z}\right)}\end{array}\right).
\end{eqnarray}

\emph{We note that all but three solutions of (\ref{sqj1}) may be written in this form.
Specifically, all solutions
of (\ref{sqj1}) bounded on the whole real line are obtained through the squared
eigenfunction connection.}

This is proven as in~\cite{bd5,bd6,bd7,bd8}: For a fixed $\lambda$, (\ref{sqj1}) is a second-order linear ordinary
differential equation, thus it has two linearly independent solutions. We know that (\ref{tsj1}) provides solutions of this ordinary differential equation. We count how many solutions are obtained in this way for a given $\lambda$. Note that $\Omega^2$ is an even function of $\zeta$. Excluding the two values of $\lambda$ for which the discriminant of (35) as a function of $\zeta$ is zero, (\ref{tsd}) gives rise to four values of $\zeta \in \mathbb{C}$. We revisit these values in (b), below. For all other values of $\lambda=2\Omega$, any fixed $\Omega$ and $\zeta$ defines a unique solution (up to a multiplicative constant) of the Lax pair.  As in~\cite{bd5,bd6,bd7,bd8}, there are two parts for this.

(a) For any $\lambda$ not equal to the two values mentioned earlier, we obtain four solutions through the squared eigenfunction connection. Since $\Omega^2$ is an even function of $\zeta$,  the Lax parameters come in $\{-\zeta, \zeta\}$ pairs. As in~\cite{bd5}, only one element of these pairs gives rise to an independent solution of the stability problem, eliminating two of these four solutions. On the other hand, as in~\cite{bd8}, if there is an exponential contribution from $\gamma$, the remaining two solutions are linearly independent. The only possibility for the exponential factor from $\gamma$ not
to contribute is if $\Omega=0$. In that case, only one linearly independent solution is obtained through the squared eigenfunction connection, corresponding to $(f_{z}, cf_{zz})^T$. The other
solution is obtained through the reduction of order method, and introduces algebraic growth.

(b) Let us consider the two excluded values of $\lambda$.
For the two $\lambda$ values where $\Omega^2$
reaches its maximum value, only one solution
is obtained through the squared eigenfunction connection, which is unbounded. The second one may be constructed using reduction of order, and introduces algebraic growth.

As a consequence of the discussion above, the double covering in
the $\Omega$ representation (\ref{tfw}) drops to a single covering. In summary, we have the following theorem.

\textbf{Theorem 2} \emph{The periodic traveling wave solutions of
the sinh-Gordon equation are spectrally stable. The spectrum of their associated linear stability problem is explicitly given by $\sigma(J \mathcal{L})=i \mathbb{R}$, or,
accounting for multiple coverings,
\begin{eqnarray}
\sigma(J \mathcal{L})=(i \mathbb{R})^2.
\end{eqnarray}}

An application of the SCS basis lemma in~\cite{by13} establishes that the eigenfunctions form a
basis for $L_{p e r}^{2}([-N \frac{T}{2}, N \frac{T}{2}])$, for any integer $N$, when the potential $f$ is periodic in $z$ with
period $T$. This
allows us to conclude linear stability with respect to subharmonic perturbations.
\\
\\
\noindent\textbf{\large 6. Orbital stability}\\\hspace*{\parindent}
In this section, we study the nonlinear stability of the elliptic solutions. We are concerned with the stability of $T$-periodic solutions of equation with respect to subharmonic perturbations of period $NT$, for any fixed positive integer $N$. We require $u$ and its derivatives of up to order three to be square-integrable.

To prove orbital stability, we prove formal stability first: we construct a Lyapunov functional for the elliptic solutions using the conserved quantities of the sinh-Gordon equation. We introduce the Hamiltonian structure of the sinh-Gordon equation and its hierarchy.
We use the system form of the sinh-Gordon equation:
\begin{eqnarray}
u_{\tau}&=&-p+cu_{z},\label{e3}\\
p_{\tau}&=&cp_{z}-u_{zz}+\sinh (u).\label{e31}
\end{eqnarray}
The Hamiltonian structure is~\cite{sinh1,sinh2}
\begin{equation}
\left(\begin{array}{c}{\partial u / \partial \tau} \\ {\partial p / \partial \tau}\end{array}\right)=J\left(\begin{array}{l}{\delta H / \delta u} \\ {\delta H / \delta p}\end{array}\right),
\end{equation}
with
\begin{equation}\label{sd}
H=-\int_{-N \frac{T}{2}}^{N \frac{T}{2}}\left[\frac{1}{2} p^{2}+\frac{1}{2}\left(u_{z}\right)^{2}+\cosh u-cpu_{z}\right] d z,
\end{equation}
and $J=\left(\begin{array}{cc}{0} & {1} \\ {-1} & {0}\end{array}\right)$.

The Hamiltonian is one of an infinite number of conserved quantities of the sinh-Gordon equation. We label these quantities $\left\{H_{j}\right\}_{j=0}^{\infty}$. We need the first three conserved quantities:
\begin{eqnarray}
&& H_{0}=-\int_{-N \frac{T}{2}}^{N \frac{T}{2}} p u_{z} d z,\nonumber\\
&& H_{1}=-\int_{-N \frac{T}{2}}^{N \frac{T}{2}}\left[\frac{1}{2} p^{2}+\frac{1}{2}\left(u_{z}\right)^{2}+\cosh u\right] d z,\nonumber\\
&& H_{2}=\int_{-N \frac{T}{2}}^{N \frac{T}{2}}\left[-64 u_{zz} p_{z}-8\left(u_{z}\right)^{3} p-8u_{z} p^{3}+48 p_{z} \sinh u\right] d z\nonumber.
\end{eqnarray}
In fact, all the functionals $H_{i}$ are mutually in involution under the Poisson
bracket. The conservation and the involution properties for $H_{0}$, $H_{1}$, and $H_{2}$ are straightforward to verify.
Each $H_{i}$ defines an evolution
equation with respect to a time variable $\tau_{i}$ by
\begin{equation}\label{gh1}
\frac{\partial}{\partial \tau_{i}}\left(\begin{array}{l}{u} \\ {p}\end{array}\right)=J H_{i}^{\prime}(p, u),
\end{equation}
where $H_{i}^{\prime}(u,p)$ denotes the variational gradient of $H_{i}$.
The collection of (\ref{gh1}) with $i=0, 1, \ldots$ is the sinh-Gordon hierarchy.  Its first three members are
\begin{eqnarray}
&& \frac{\partial}{\partial \tau_{0}}\left(\begin{array}{l}{u} \\ {p}\end{array}\right)=\left(\begin{array}{l}{-u_{z}} \\ {-p_{z}}\end{array}\right),\nonumber\\
&& \frac{\partial}{\partial \tau_{1}}\left(\begin{array}{l}{u} \\ {p}\end{array}\right)=\left(\begin{array}{l}{-p} \\ {\sinh u-u_{zz}}\end{array}\right),\nonumber\\
&& \frac{\partial}{\partial \tau_{2}}\left(\begin{array}{l}{u} \\ {p}\end{array}\right)=\left(\begin{array}{l}-8{u_{z}}^3-24u_{z}p^2-48u_{z}\cosh u+64u_{zzz} \\ -48p_{z}\cosh u-24p^2p_{z}-24p_{z}{u_{z}}^2-48pu_{zz}u_{z} +64p_{zzz} \end{array}\right).\nonumber
\end{eqnarray}
Since the flows in the sinh-Gordon hierarchy commute, we may take any linear combination of the above  Hamiltonians to define a new Hamiltonian system. We define the
$n$-th sinh-Gordon equation with evolution variable $t_{n}$ as
\begin{eqnarray}
&&\frac{\partial}{\partial t_{n}}\left(\begin{array}{l}{u} \\ {p}\end{array}\right)=J \hat{H}_{n}^{\prime}(u, p),\\
&& \hat{H}_{n}=H_{n}+\sum_{i=0}^{n-1} c_{n, i} H_{i}, n \geqslant 1,
\end{eqnarray}
where the coefficients $c_{n,i}$ are constants. For example, $\hat{H}_{1}=H_{1}-cH_{0}=H$ is the Hamiltonian of the sinh-Gordon equation~(\ref{e3}) and~(\ref{e31}), as shown in~(\ref{sd}).

Every member of the sinh-Gordon hierarchy has a Lax pair. These Lax pairs share the same first component $-T_{0}$, while the second component $\psi_{\tau_{j}}=T_{j} \psi$ is different:
\begin{eqnarray}
&& \psi_{\tau_{0}}=T_{0}\psi=-\left(\begin{array}{cc}{\frac{i \cosh u}{8 \zeta }-i \frac{\zeta}{2}} & {\frac{1}{4} \left(u_{z}-p\right)-\frac{i \sinh u}{8 \zeta }} \\ {\frac{1}{4} \left(u_{z}-p\right)+\frac{i \sinh u}{8 \zeta }} & -{{\frac{i \cosh u}{8 \zeta }+i \frac{\zeta}{2}}}\end{array}\right) \psi,\nonumber\\
&& \psi_{\tau_{1}}=T_{1}\psi=\left(\begin{array}{cc}{-\frac{i \cosh u}{8 \zeta }-i \frac{\zeta}{2}} & {\frac{1}{4} \left(u_{z}-p\right)+\frac{i \sinh u}{8 \zeta }} \\ {\frac{1}{4} \left(u_{z}-p\right)-\frac{i \sinh u}{8 \zeta }} & {{\frac{i \cosh u}{8 \zeta }+i \frac{\zeta}{2}}}\end{array}\right) \psi,\nonumber\\
&& \psi_{\tau_{2}}=T_{2}\psi=\left(\begin{array}{cc}{A_{2}} & {B_{2}} \\ {C_{2}} & {-A_{2}}\end{array}\right) \psi,\nonumber
\end{eqnarray}
where
\begin{eqnarray}
A_{2}&=&32 i \zeta ^3-\frac{i \cosh u}{2 \zeta^3}+8 i \zeta  \left(-p u_{z}+\frac{1}{2} p^2+\frac{1}{2} u_{z}^2\right)\nonumber\\
&&+\frac{2}{\zeta }\left(-2 i p_{z} \text{csch}u+2 i p_{z} \cosh u \coth u-i p u_{z} \cosh u-\frac{1}{2} i p^2 \cosh u-\frac{1}{2} i u_{z}^2 \cosh u\right.\nonumber\\
&&\left.-2 i u_{zz} \text{csch}u+2 i u_{zz} \cosh u \coth u-i \cosh ^2 u+i\right),\nonumber\\
B_{2}&=&\frac{i \sinh u}{2 \zeta ^3}+\frac{1}{16} i \zeta ^2 \left(256 i u_{z}-256 i p\right)+\frac{ p+ u_{z}}{\zeta ^2}+8 i\zeta  \left(2 p_{z}-2 u_{zz}+\sinh u\right)\nonumber\\
&&+\frac{i \sinh u}{ \zeta }(-4 p_{z} \coth u+2 p u_{z}+p^2+u_{z}^2-4 u_{zz} \coth u+2 \cosh u)+\frac{i}{128}(-512 i p \cosh u\nonumber\\
&&-256 i p^3+2048 i p_{zz}-768 i p u_{z}^2+768 i p^2 u_{z}+256 i u_{z}^3-2048 i u_{zzz}+1536 i u_{z} \cosh u),\nonumber\\
C_{2}&=&\frac{-i \sinh u}{2 \zeta ^3}+\frac{1}{16} i \zeta ^2 \left(256 i u_{z}-256 i p\right)+\frac{ p+ u_{z}}{\zeta ^2}-8 i \zeta  \left(2 p_{z}-2 u_{zz}+\sinh u\right)\nonumber\\
&&-\frac{i \sinh u}{\zeta }(-4 p_{z} \coth u+2 p u_{z}+p^2+u_{z}^2-4 u_{zz} \coth u+2 \cosh u)+\frac{i}{128}(-512 i p \cosh u\nonumber\\
&&-256 i p^3+2048 i p_{zz}-768 i p u_{z}^2+768 i p^2 u_{z}+256 i u_{z}^3-2048 i u_{zzz}+1536 i u_{z} \cosh u).\nonumber
\end{eqnarray}
We construct the Lax pair for the $n$-th
sinh-Gordon equation by taking the same linear combination of the lower-order flows as for the nonlinear hierarchy:
\begin{eqnarray}
\psi_{t_{n}}&=&\hat{T}_{n} \psi=\left(\begin{array}{cc}{\hat{A}_{n}} & {\hat{B}_{n}} \\ {\hat{C}_{n}} & {-\hat{A}_{n}}\end{array}\right) \psi, \\ \hat{T}_{n} &=&T_{n}+\sum_{i=0}^{n-1} c_{n, i} T_{i}, \quad \hat{T}_{0} =T_{0}.
\end{eqnarray}

We study the stationary solutions of the sinh-Gordon hierarchy. Since the flows commute, a stationary solution of the $n$th equation remains a stationary solution after evolving under any of the other
flows, for a suitable choice of the coefficients $c_{n,i}$.

For example, the traveling wave solutions $(f, c f_{z})$  are the stationary solutions of the first equation in the sinh-Gordon hierarchy with $c_{1,0}=-c$.  They are also stationary solutions of the second equation in the sinh-Gordon hierarchy provided
\begin{eqnarray}
\frac{16 \left(-3 c^2-1\right) \mathcal{E}}{\left(c^2-1\right)}-c_{2,1}c-c_{2,0}=0.
\end{eqnarray}

This gives one condition for the two coefficients $c_{2,1}$ and $c_{2,0}$.
In order to proceed as in Refs.~\cite{by14,bd6}, we consider stability in the space of subharmonic functions of period $NT$, for $1\leq N\in \mathbb{N}$, i.e.,
\begin{equation}
\mathbb{V}_{0,N}=\left\{W: W \in H_{p e r}^{3}([-N \frac{T}{2}, N \frac{T}{2}]) \right\}.
\end{equation}

To prove the orbital stability of the solution $(u, p)$ in this space, we construct a Lyapunov function~\cite{by14,by15}, i.e., a constant of the motion $\mathcal{E}(u,p)$ for which $(u,p)$ is an unconstrained minimizer:

\begin{equation}\label{tc1}
\frac{d \mathcal{E}(u,p)}{d \tau}=0, \quad \mathcal{E}^{\prime}(u,p)=0, \quad\left\langle v, {\cal L}(u,p) v\right\rangle> 0,\quad \forall v \in \mathbb{V}_{0},\quad  v \neq 0,
\end{equation}
where $\mathcal{E}^{\prime}(u,p)$ denotes the variational gradient of $\mathcal{E}$ and $\cal L$ is the Hessian of $\mathcal{E}$.
The existence of such a function implies formal stability. We know that $(f_{z}, c f_{zz})^{T}$ is in the kernel of $H_{1}^{\prime \prime}=\mathcal{L}$. This is obtained from the action of the
infinitesimal generator $\partial_{z}$, acting on $(f(z), p(z))^{T}$, where $p(z)=cf_{z}$.
Following Grillakis, Shatah, and Strauss~\cite{lf4,by14}, under extra conditions (see the orbital stability theorem in~\cite{bd4,bd5,bd6,bd7}) this allows one to conclude orbital stability.
Since the sinh-Gordon equation is an integrable Hamiltonian system, all the conserved quantities of the equation satisfy the first two conditions. It suffices to construct one that satisfies the third requirement.

To prove orbital stability,
we check the Krein signature $K_{1}$~\cite{lf4}, associated with $\hat{H}_{1}$:
\begin{equation}
K_{1}=\left\langle W, \mathcal{L}_{1} W\right\rangle=\int_{-N \frac{T}{2}}^{N  \frac{T}{2}} W^{*} \mathcal{L}_{1} W d z,
\end{equation}
where $\mathcal{L}_{1}=\mathcal{L}$.
Using the squared eigenfunction connection, we have
\begin{eqnarray}
W^{*} \mathcal{L}_{1} W&=&2 \Omega W^{*} J^{-1} W\nonumber\\
&=&2 \Omega\left(W_{1} W_{2}^{*}-W_{2} W_{1}^{*}\right)\nonumber\\
&=&8\Omega^2(\left|\varphi_{1}\right|^{4}+\left|\varphi_{2}\right|^{4}-\varphi_{1}^2{\varphi^{*}_{2}}^2-\varphi_{2}^2{\varphi^{*}_{1}}^2)+2\Omega(2c\varphi^2_{1}\varphi^{*}_{1}\varphi^{*}_{1z}+2c\varphi^2_{2}\varphi^{*}_{2}\varphi^{*}_{2z}-2c\varphi^2_{2}\varphi^{*}_{1}\varphi^{*}_{1z}\nonumber\\
&&-2c\varphi^2_{1}\varphi^{*}_{2}\varphi^{*}_{2z}-2c\varphi^{* 2}_{1}\varphi_{1}\varphi_{1z}-2c\varphi^{* 2}_{2}\varphi_{2}\varphi_{2z} +2c\varphi^{* 2}_{1}\varphi_{2}\varphi_{2z}+2c\varphi^{* 2}_{2}\varphi_{1}\varphi_{1z}),
\end{eqnarray}
with $\varphi_{1}=-\gamma(z) B(z)$ and $\varphi_{2}=\gamma(z) (A(z)-\Omega)$. Using (\ref{sim1}), we have
\begin{eqnarray}
|\gamma|^2=\frac{1}{|A-\Omega|}.
\end{eqnarray}
With $\Omega^2=A^2+|B|^2$, we have
\begin{eqnarray}
&&|\varphi_{1}|^4=-(A+\Omega)^2,\quad |\varphi_{2}|^4=-(A-\Omega)^2,\quad \varphi_{2}^2{\varphi^{*}_{1}}^2=-{B^{*}}^{2},\quad \varphi_{1}^2{\varphi^{*}_{2}}^2=-{B}^{2},\nonumber\\
&&\varphi^2_{1}\varphi^{*}_{1}\varphi^{*}_{1z}-\varphi^{* 2}_{1}\varphi_{1}\varphi_{1z}=\frac{A+\Omega}{A-\Omega}(BB^{*}_{z}-B^{*}B_{z})-(\Omega+A)^2(\Theta^{*}-\Theta),\nonumber\\ &&\varphi^2_{2}\varphi^{*}_{2}\varphi^{*}_{2z}-\varphi^{* 2}_{2}\varphi_{2}\varphi_{2z}=-(A-\Omega)^2(\Theta^{*}-\Theta),\nonumber\\
&&\varphi^{* 2}_{1}\varphi_{2}\varphi_{2z}-\varphi^2_{2}\varphi^{*}_{1}\varphi^{*}_{1z}=B^{* 2}(\Theta^{*}-\Theta)+B^{*}B^{*}_{z}-\frac{B^{* 2}}{A-\Omega}A_{z},\nonumber\\
&&\varphi^{* 2}_{2}\varphi_{1}\varphi_{1z}-\varphi^2_{1}\varphi^{*}_{2}\varphi^{*}_{2z}=B^{2}(\Theta^{*}-\Theta)-BB_{z}+\frac{B^2}{A-\Omega}A_{z},\nonumber
\end{eqnarray}
where
\begin{eqnarray}
\Theta=\frac{i}{2} \zeta-\frac{i\cosh f(z)}{8\zeta} - \frac{B \left[{\frac{(1-c)f'(z)}{4}+\frac{i\sinh f(z)}{8\zeta}}\right]+A_{z}}{A-\Omega}.
\end{eqnarray}
It follows that the Krein signature $K_{1}$ can be expressed as
\begin{eqnarray}\label{ks}
K_{1}&=&\left\langle W, \mathcal{L}_{1} W\right\rangle=-\Omega^2\int_{-N \frac{K}{b}}^{N  \frac{K}{b}} \left(\frac{-16 \zeta^4 (c+1)-8 \zeta^2 \cosh (f(z))+c-1}{2 \zeta^2}\right) d z\nonumber\\
&=&-\frac{N\Omega^2}{b}\left[\left(\frac{-1+c}{\zeta^2}-16\zeta^2(1+c)+8\right)K(k)-16\frac{E(k)}{1-k^2}\right]\\
&=&-\frac{N\Omega^2 P(\zeta)}{b \zeta^2},
\end{eqnarray}
where $E(k)$ is the complete elliptic integral of the second kind~\cite{rb1}:
\begin{eqnarray}
E(k)=\int_{0}^{\pi / 2} \sqrt{1-k^{2} \sin ^{2} y} \ \mathrm{d} y.
\end{eqnarray}

We have the following properties:

1. $P(\zeta)$ is an even function and  the discriminant of $P(\zeta)$ is positive. Since $\frac{-1+c}{-16(1+c)}<0$, $P(\zeta)=0$ has two real roots $\pm \zeta_{c}$ $(\zeta_{c}>0)$. It follows that $K_{1}(\zeta)=0$, when $\Omega(\zeta)=0$ or $\zeta=\pm \zeta_{c}$. Since $\frac{d{\frac{P(\zeta)}{\zeta^2}}}{d\zeta}=-\frac{2 \left(16 \zeta^4 (c+1)+c-1\right) K\left(\sqrt{\frac{\mathcal{E}-1}{\mathcal{E}+1}}\right)}{\zeta^3}$,
we know that for $c>1$, when $\zeta>0$, $\frac{P(\zeta)}{\zeta^2}$ decreases along $\zeta$ and when $\zeta<0$, $\frac{P(\zeta)}{\zeta^2}$ increases along $\zeta$. For $c<-1$, when $\zeta>0$, $\frac{P(\zeta)}{\zeta^2}$ increases along $\zeta$ and when $\zeta<0$, $\frac{P(\zeta)}{\zeta^2}$ decreases along $\zeta$.

2. Since $\zeta_{1}<\zeta_{c}<\zeta_{2}$, $\pm\zeta_{c}$ is not in $\sigma_L$ (see Appendix), $K_{1}=0$ is obtained only on the kernel of $\mathcal{L}_{1}$, i.e., when $\Omega=0$. For $c<-1$, $\zeta_{c}=\frac{1}{2} \sqrt{\frac{-\sqrt{c^2 K(k)^2+E^2(k) (\mathcal{E}+1)^2-2 E(k) (\mathcal{E}+1) K(k)}-E(k) (\mathcal{E}+1)+K(k)}{(c+1) K(k)}}$. For $c>1$, $\zeta_{c}=\frac{1}{2} \sqrt{\frac{\sqrt{c^2 K(k)^2+E^2(k) (\mathcal{E}+1)^2-2 E(k) (\mathcal{E}+1) K(k)}-E(k) (\mathcal{E}+1)+K(k)}{(c+1) K(k)}}$. We note that $\zeta_{c}, \zeta_{1}$ and $\zeta_{2}$ are all greater than zero.

3. When $c>1$, since $\frac{P(\zeta)}{\zeta^2}>0$ for $|\zeta|< \zeta_{1}$, $K_{1}>0$ for $|\zeta|< \zeta_{1}$ and since $\frac{P(\zeta)}{\zeta^2}<0$ for $|\zeta|> \zeta_{2}$, $K_{1}<0$ for $|\zeta|>\zeta_{2}$. When $c<-1$, since $\frac{P(\zeta)}{\zeta^2}<0$ for $|\zeta|< \zeta_{1}$, $K_{1}<0$ for $|\zeta|< \zeta_{1}$ and since $\frac{P(\zeta)}{\zeta^2}>0$ for $|\zeta|> \zeta_{2}$, $K_{1}>0$ for $|\zeta|>\zeta_{2}$. Thus no
conclusion about orbital stability can be drawn from $K_{1}$.

It follows that $\hat{H}_{1}$ is not a Lyapunov function. Thus we need to consider different conserved quantities.
Linearizing the $n$-th sinh-Gordon equation about the equilibrium solution $f$, one obtains
\begin{eqnarray}
w_{t_{n}}=J \mathcal{L}_{n} w,
\end{eqnarray}
where $\mathcal{L}_{n}$ is the Hessian of $\hat{H}_{n}$ evaluated at the stationary solution.

Using the squared-eigenfunction connection with separation of variables gives
\begin{equation}\label{ttx1}
2 \Omega_{n} W(z)=J \mathcal{L}_{n} W(z),
\end{equation}
where $\Omega_{n}$ is defined through
\begin{equation}\label{tta}
\psi\left(z, t_{n}\right)=e^{\Omega_{n} t_{n}} \varphi(z).
\end{equation}
Due to the commuting property of the different flows in the sinh-Gordon hierarchy, the Lax hierarchy shares the
common set of eigenfunctions $\varphi(z)$ from before (still assuming the
solution is stationary with respect to the first, and hence all higher flows). Substituting~(\ref{tta}) into the Lax pair of the $n$-th sinh-Gordon equation determines a relationship between $\Omega_{n}$ and $\zeta$
\begin{equation}
\Omega_{n}^{2}(\zeta)=\hat{A}_{n}^{2}+\hat{B}_{n} \hat{C}_{n}.
\end{equation}
To find a Lyapunov functional,
we check $K_{2}$:
\begin{equation}
K_{2}=\int_{-N  \frac{T}{2}}^{N  \frac{T}{2}} W^{*} \mathcal{L}_{2} W d z=2 \Omega_{2} \int_{-N  \frac{T}{2}}^{N  \frac{T}{2}} W^{*} J^{-1} W d z=\frac{\Omega_{2}}{\Omega}\int_{-N  \frac{T}{2}}^{N  \frac{T}{2}} W^{*} \mathcal{L}_{1} W d z.
\end{equation}
Therefore, we have
\begin{equation}
K_{2}(\zeta)=\Omega_{2}(\zeta) \frac{K_{1}(\zeta)}{\Omega(\zeta)}.
\end{equation}
Here, we use that $(f, c f_{z})$ are the stationary solutions of the second flow.
To calculate $K_{2}$, we also need
\begin{equation}
\hat{T}_{2}=T_{2}+c_{2,1} T_{1}+c_{2,0} T_{0},
\end{equation}
where, from before,
\begin{eqnarray}\label{yy1}
\frac{16 \left(-3 c^2-1\right)\mathcal{E}}{\left(c^2-1\right)}-c_{2,1}c-c_{2,0}=0.
\end{eqnarray}
The second sinh-Gordon equation can be expressed as
\begin{equation}
\frac{\partial}{\partial t_{2}}\left(\begin{array}{l}{u} \\ {p}\end{array}\right)=J\left(H_{2}^{\prime}+c_{2,1} H_{1}^{\prime}+c_{2,0} H_{0}^{\prime}\right)=0.
\end{equation}
A direct calculation gives
\begin{eqnarray}
\Omega^2_{2}=\frac{\left[c^2 \left(64 \zeta ^4+16 \zeta ^2 \mathcal{E}+\zeta ^2 c_{2,0}+4\right)+c \left(4-64 \zeta ^4\right)+\zeta ^2 (16 \mathcal{E}-c_{2,0})\right]^2}{c^2 \left(c^2-1\right)^2 \zeta ^4}\Omega^2.
\end{eqnarray}
We can choose
\begin{eqnarray}
c_{2,0}=-\frac{4 \left(16 c^2 \zeta_{c} ^4+4 c^2 \zeta_{c} ^2 \mathcal{E}+c^2-16 c \zeta_{c} ^4+c+4 \zeta_{c} ^2 \mathcal{E}\right)}{\left(c^2-1\right) \zeta_{c} ^2},
\end{eqnarray}
to ensure $K_{2}$ has definite sign. With this choice of $c_{2,0}$ and $c_{2,1}$ determined by~(\ref{yy1}), $\hat{H}_{2}$ is a Lyapunov functional for the dynamics (with respect to any of the time variables in the sinh-Gordon hierarchy) of the stationary solutions.
Therefore, whenever solutions are spectrally
stable with respect to subharmonic perturbations of period $N$, they are formally stable in $\mathbb{V}_{0,N}$.

Since the
infinitesimal generators of the symmetries correspond to the values of $\zeta$ for which $\Omega(\zeta)=0$, the kernel of the functional $\hat{H}_{2}^{\prime \prime}(u, p)$ consists of the infinitesimal generators of the symmetries of the solution $(u,p)$. On the other hand, since $\pm\zeta_{c}$ is not in $\sigma_L$, $K_{2}(\zeta)=0$ is obtained only when $\Omega=0$ for $\zeta\in\sigma_L$.
We have proved the following theorem.
\\
\\
\textbf{Theorem 3} \textbf{(Orbital stability)}
The elliptic solutions~(\ref{sc1}) of the sinh-Gordon equation are
orbitally stable with respect to subharmonic perturbations in $\mathbb{V}_{0,N}, N\geq 1$.
\\
\\
\noindent\textbf {Acknowledgments}
WS has been supported by the National Natural Science Foundation of China under Grant No.61705006, and by the Fundamental Research Funds of the Central Universities (No.230201606500048).\\

\noindent\textbf {\large Appendix}\\
\textbf{Lemma.} \emph{For $c>1$, $\frac{P(\zeta_{1})}{\zeta^2_{1}}>0$ and $\frac{P(\zeta_{2})}{\zeta^2_{2}}<0$, while for $c<-1$, $\frac{P(\zeta_{1})}{\zeta^2_{1}}<0$ and $\frac{P(\zeta_{2})}{\zeta^2_{2}}>0$.}\\
Proof. \\
$\bullet$ For $c>1$,
\begin{eqnarray}
\frac{P(\zeta_{1})}{\zeta^2_{1}}=8 c \sqrt{\mathcal{E}^2-1} K\left(\sqrt{\frac{\mathcal{E}-1}{\mathcal{E}+1}}\right)+8 (\mathcal{E}+1) K\left(\sqrt{\frac{\mathcal{E}-1}{\mathcal{E}+1}}\right)-8 (\mathcal{E}+1) E\left(\sqrt{\frac{\mathcal{E}-1}{\mathcal{E}+1}}\right)
.
\end{eqnarray}
Since $E(k)<K(k)$, $c>1$ and $\mathcal{E}>1$, we have $\frac{P(\zeta_{1})}{\zeta^2_{1}}>0$.
\begin{eqnarray}
\frac{P(\zeta_{2})}{\zeta^2_{2}}=
8 c \left(-\sqrt{\mathcal{E}^2-1}\right) K\left(\sqrt{\frac{\mathcal{E}-1}{\mathcal{E}+1}}\right)+8 (\mathcal{E}+1) K\left(\sqrt{\frac{\mathcal{E}-1}{\mathcal{E}+1}}\right)-8 (\mathcal{E}+1) E\left(\sqrt{\frac{\mathcal{E}-1}{\mathcal{E}+1}}\right).
\end{eqnarray}
Let $\frac{P(\zeta_{2})}{\zeta^2_{2}}=F(c)$. We note that $F'(c)=8\left(-\sqrt{\mathcal{E}^2-1}\right) K\left(\sqrt{\frac{\mathcal{E}-1}{\mathcal{E}+1}}\right)<0$. We have
\begin{eqnarray}
F(c)<F(1)=8 \left(-\sqrt{\mathcal{E}^2-1}\right) K\left(\sqrt{\frac{\mathcal{E}-1}{\mathcal{E}+1}}\right)+8 (\mathcal{E}+1) K\left(\sqrt{\frac{\mathcal{E}-1}{\mathcal{E}+1}}\right)-8 (\mathcal{E}+1) E\left(\sqrt{\frac{\mathcal{E}-1}{\mathcal{E}+1}}\right).
\end{eqnarray}
Using $\frac{E(k)}{K(k)}>k'=\sqrt{1-k^2}$ , see [1, 19.9.8], we have
\begin{eqnarray}
8 \left(-\sqrt{\mathcal{E}^2-1}\right)+8 (\mathcal{E}+1)-8 (\mathcal{E}+1) \frac{E\left(\sqrt{\frac{\mathcal{E}-1}{\mathcal{E}+1}}\right)}{K\left(\sqrt{\frac{\mathcal{E}-1}{\mathcal{E}+1}}\right)}<8 \left(-\sqrt{\mathcal{E}^2-1}\right)+8 (\mathcal{E}+1)-8\sqrt{2}\sqrt{\mathcal{E}+1}.
\end{eqnarray}
Let $Q(\mathcal{E})=8 \left(-\sqrt{\mathcal{E}^2-1}\right)+8 (\mathcal{E}+1)-8\sqrt{2}\sqrt{\mathcal{E}+1}$. We note $Q'(\mathcal{E})=-\frac{8 \mathcal{E}}{\sqrt{\mathcal{E}^2-1}}+8-\frac{4 \sqrt{2}}{\sqrt{\mathcal{E}+1}}<-\frac{4 \sqrt{2}}{\sqrt{\mathcal{E}+1}}<0$. So we have
$Q(\mathcal{E})<Q(1)=0$ for $\mathcal{E}>1$. Therefore, we have $\frac{P(\zeta_{2})}{\zeta^2_{2}}=F(c)<F(1)<K\left(\sqrt{\frac{\mathcal{E}-1}{\mathcal{E}+1}}\right)Q(\mathcal{E})<0$.\\
$\bullet$ For $c<-1$,
\begin{eqnarray}
\frac{P(\zeta_{1})}{\zeta^2_{1}}=8 c \sqrt{\mathcal{E}^2-1} K\left(\sqrt{\frac{\mathcal{E}-1}{\mathcal{E}+1}}\right)+8 (\mathcal{E}+1) K\left(\sqrt{\frac{\mathcal{E}-1}{\mathcal{E}+1}}\right)-8 (\mathcal{E}+1) E\left(\sqrt{\frac{\mathcal{E}-1}{\mathcal{E}+1}}\right).
\end{eqnarray}
Let $\frac{P(\zeta_{1})}{\zeta^2_{1}}=G(c)$. We note that $G'(c)=8\left(\sqrt{\mathcal{E}^2-1}\right) K\left(\sqrt{\frac{\mathcal{E}-1}{\mathcal{E}+1}}\right)>0$. We have
\begin{eqnarray}
G(c)<G(-1)=8 \left(-\sqrt{\mathcal{E}^2-1}\right) K\left(\sqrt{\frac{\mathcal{E}-1}{\mathcal{E}+1}}\right)+8 (\mathcal{E}+1) K\left(\sqrt{\frac{\mathcal{E}-1}{\mathcal{E}+1}}\right)-8 (\mathcal{E}+1) E\left(\sqrt{\frac{\mathcal{E}-1}{\mathcal{E}+1}}\right).
\end{eqnarray}
Again, using $\frac{E(k)}{K(k)}>k'=\sqrt{1-k^2}$, we have
\begin{eqnarray}
8 \left(-\sqrt{\mathcal{E}^2-1}\right)+8 (\mathcal{E}+1)-8 (\mathcal{E}+1) \frac{E\left(\sqrt{\frac{\mathcal{E}-1}{\mathcal{E}+1}}\right)}{K\left(\sqrt{\frac{\mathcal{E}-1}{\mathcal{E}+1}}\right)}<8 \left(-\sqrt{\mathcal{E}^2-1}\right)+8 (\mathcal{E}+1)-8\sqrt{2}\sqrt{\mathcal{E}+1}.
\end{eqnarray}
We know
$Q(\mathcal{E})<Q(1)=0$ for $\mathcal{E}>1$. Therefore, $\frac{P(\zeta_{1})}{\zeta^2_{1}}=G(c)<G(-1)<K\left(\sqrt{\frac{\mathcal{E}-1}{\mathcal{E}+1}}\right)Q(\mathcal{E})<0$.
\begin{eqnarray}
\frac{P(\zeta_{2})}{\zeta^2_{2}}=
8 c \left(-\sqrt{\mathcal{E}^2-1}\right) K\left(\sqrt{\frac{\mathcal{E}-1}{\mathcal{E}+1}}\right)+8 (\mathcal{E}+1) K\left(\sqrt{\frac{\mathcal{E}-1}{\mathcal{E}+1}}\right)-8 (\mathcal{E}+1) E\left(\sqrt{\frac{\mathcal{E}-1}{\mathcal{E}+1}}\right).
\end{eqnarray}
Since $E(k)<K(k)$, $c<-1$ and $\mathcal{E}>1$, we have $\frac{P(\zeta_{2})}{\zeta^2_{2}}>0$. This finishes the proof of the Lemma.


\begin{thebibliography}{99}
\bibitem{jk1}
NIST Digital Library of Mathematical Functions. http://dlmf.nist.gov/, Release 1.0.14 of 2016-12-21. F. W. J. Olver,
A. B. Olde Daalhuis, D. W. Lozier, B. I. Schneider, R. F. Boisvert, C. W. Clark, B. R. Miller and B. V. Saunders, eds.
\bibitem{sinh1}
Ablowitz M J, Segur H. Solitons and the Inverse Scattering Transform, SIAM,
Philadelphia, PA, 1981.
\bibitem{tt11}
Arnold V I. Mathematical methods of classical mechanics. Springer-Verlag, New York, NY,
1997.
\bibitem{tt22}
Arnold V I. On an a priori estimate in the theory of hydrodynamical stability. Am. Math.
Soc. Transl. 1969, 79, 267-269.
\bibitem{bd6}
Bottman N, Deconinck B. Nivala M. Elliptic solutions of the defocusing NLS equation are stable. J. Phys. A  2011, 44, 285201.
\bibitem{bd8}
Bottman N, Deconinck B. KdV cnoidal waves are spectrally stable. DCDS-A 2009,  25, 1163-1180.
\bibitem{sinh3}
Chern S S. Geometrical interpretation of the sinh-Gordon equation. Ann. Pol. Math. 1981, 1, 63-69.
\bibitem{bd4}
Deconinck B, Kapitula T. The orbital stability of the cnoidal waves of the Korteweg-de Vries equation. Phys. Lett. A 2010, 374, 4018-4022.
\bibitem{bd5}
Deconinck B, Nivala M. The stability analysis of the periodic traveling wave solutions of the mKdV equation. Stud. Appl. Math. 2011, 126, 17-48.
\bibitem{t1}
Deconinck B, Segal B L. The stability spectrum for elliptic solutions to the focusing NLS equation. Physica D 2017, 346, 1-19.
\bibitem{by90}
Deconinck B, Upsal J. The orbital stability of elliptic solutions of the Focusing Nonlinear Schr\"{o}dinger Equation. SIAM J. Math. Anal. 2020, 52, 1-41.
\bibitem{ees1}
Deconinck B, McGill P, Segal B L. The stability spectrum for elliptic solutions to the sine-Gordon equation. Physica D 2017, 360, 17-35.
\bibitem{by121}
Deconinck B, Kapitula T. On the orbital (in) stability of spatially periodic stationary solutions of generalized Korteweg-de Vries equations. preprint, 2010.
\bibitem{lf4}
Grillakis M, Shatah J, Strauss W. Stability theory of solitary waves in the presence of symmetry. I J. Funct. Anal. 1987, 74, 160-197.
\bibitem{by14}
Grillakis M, Shatah J, Strauss W. Stability theory of solitary waves in the presence of symmetry, II. J. Funct. Anal. 1990, 94, 308-348.
\bibitem{lf2}
Holm D D, Marsden J E, Ratiu T, Weinstein A. Nonlinear stability of fluid and plasma equilibria. Phys. Rep. 1985, 123, 1-116.
\bibitem{lf3}
Henry D B, Perez J F, Wreszinski W F. Stability theory for solitary-wave solutions of scalar field equations. Comm. Math. Phys. 1982, 85, 351-361.
\bibitem{by13}
Haragus M, Kapitula T. On the spectra of periodic waves for infinite-dimensional Hamiltonian systems. Physica D 2008, 237, 2649-2671.
\bibitem{ees2}
Jones C K R T, Marangell R, Miller P D, Plaza R G. On the stability analysis of periodic sine-Gordon traveling waves. Physica D 2013, 251, 63-74.
\bibitem{sinh4}
Larsen A L, Sanchez N. sinh-Gordon, cosh-Gordon, and Liouville equations for strings and multistrings in constant curvature spacetimes. Phys. Rev. D 1996, 54, 2801-2807.
\bibitem{rb1}
Lawden D F. Elliptic Functions and Applications (Applied Mathematical Sciences vol
80), New York, Springer, 1989.
\bibitem{sinh2}
McKean H P. The sin-gordon and sinh-gordon equations on the circle. Comm. Pur. Appl. Math. 1981, 34, 197-257.
\bibitem{by15}
Maddocks J H, Sachs R L. On the stability of KdV multi-solitons. Comm. Pur.  Appl. Math. 1993, 46, 867-901.
\bibitem{sinh5}
Natali F. On periodic waves for sine-and sinh-Gordon equations. J. Math. Anal. Appl. 2011, 379, 334-350.
\bibitem{bd7}
Nivala M, Deconinck B. Periodic finite-genus solutions of the KdV equation are orbitally
stable. Physica D 2010, 239, 1147-1158.
\bibitem{by12}
Newell A C. Solitons in Mathematics and Physics, Vol. 48, SIAM, Philadelphia, PA,
1985.
\bibitem{by11}
Sachs R L. Completeness of derivatives of squared Schr\"{o}dinger eigenfunctions and explicit solutions of the linearized KdV equation. SIAM J. Math. Anal. 1983, 14, 674-683.
\bibitem{jm}
Upsal J, Deconinck B. Real Lax spectrum implies spectral stability. arXiv preprint arXiv:1909.10119v2,2019.
\bibitem{lf1}
Weinstein M I. Lyapunov stability of ground states of nonlinear dispersive evolution equations. Comm. Pure Appl. Math. 1986, 39, 51-67.
\bibitem{bbby9}
Wiggins S. Introduction to applied nonlinear dynamical systems and chaos, volume 2.
Springer-Verlag, New York, second edition, 2003.
\end{thebibliography}
\end{document}